\documentclass{article}
\usepackage{spconf,amsmath,amssymb,graphicx}

\usepackage{mathtools}
\usepackage{multirow}
\usepackage{adjustbox}
\usepackage{xcolor}
\usepackage{subcaption}
\usepackage{hyperref}
\usepackage{enumitem}
\usepackage{algorithm}
\usepackage{algorithmic}
\usepackage{tablefootnote}
\usepackage{booktabs}
\usepackage{tikz}
\tikzstyle{arrow} = [thick,->,>=stealth,line width=0.5pt]
\usepackage{color}

\usepackage{array,mathtools}
\newcolumntype{P}[1]{>{\centering\arraybackslash}p{#1}}

\title{Inference and Denoise: Causal Inference-based \\Neural Speech Enhancement}
\name{Tsun-An Hsieh$^{1}$, Chao-Han Huck Yang$^{2}$, Pin-Yu Chen$^{3}$,
      Sabato Marco Siniscalchi$^{2,4}$,
      Yu Tsao$^{1}$
}

\address{
$^1$Research Center for Information Technology Innovation, Academia Sinica, Taiwan \\
$^2$Georgia Institute of Technology, GA, USA; $^3$IBM Research, NY, USA\\
$^4$Computer Engineering School, Norwegian University of Science and Technology, Norway
}
\begin{document}
\ninept
\maketitle
\begin{abstract}
This study addresses the speech enhancement (SE) task within the causal inference paradigm by modeling the noise presence as an intervention. Based on the potential outcome framework, the proposed causal inference-based speech enhancement (CISE) separates clean and noisy frames in an intervened noisy speech using a noise detector and assigns both sets of frames to two mask-based enhancement modules (EMs) to perform noise-conditional SE. Specifically, we use the presence of noise as guidance for EM selection during training, and the noise detector selects the enhancement module according to the prediction of the presence of noise for each frame. Moreover, we derived a SE-specific average treatment effect to quantify the causal effect adequately. Experimental evidence demonstrates that CISE outperforms a non-causal mask-based SE approach in the studied settings and has better performance and efficiency than more complex SE models.
\end{abstract}
\section{Introduction}
\label{sec:intro}
Recent advances in neural network-based speech enhancement (SE) have demonstrated impressive performance in terms of speech quality and intelligibility scores, such as perceptual evaluation of speech quality (PESQ) \cite{pesq} and short-time objective intelligibility (STOI) \cite{taal2010short} in various speech applications.
However, modern SE approaches \cite{lu2013speech, xu2014regression, Wang2018, Qi2020slp, hartmann2013direct} do not explicitly take the presence of noise into account.
, and real-world acoustic scenarios often encounter inevitable observational uncertainties, For instance, a meeting could be abruptly disconnected, or a session could be disrupted by temporary noise from the external environment. That is, noise intervention may not affect the entire speech waveform. In such a scenario, conventional neural SE solutions %
may be unreliable for handling the intermittent/sporadic nature of the noise. By contrast, causal inference (CI) \cite{pearl2010causal} may be a viable paradigm for performing SE. The design of an end-to-end neural SE model within the CI framework is the research question addressed in this study. 
Causal inference-based machine learning  techniques are often featured with the ability to identify unobserved factors or features (also know as confounding variables) with improved model prediction and generalization, i.e., CI-based models are proven to be advantageous of tackling unseen data hence dependable~\cite{louizos2017causal, johansson2016learning}. Furthermore, machine learning models that satisfy the CI training objectives stand to benefit from additional interpretable scores to formally quantify the causal effects, for example, in treatment effect estimation. Previous studies~\cite{athey2015machine, tang2021model} have demonstrated that learning to measure causal variables empowers effective model selection.
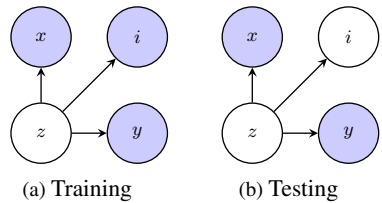
\begin{figure}[t!]
	\centering
	\begin{subfigure}{0.10\textwidth} %
	    \centering
	\begin{tikzpicture}[x = 2.0cm, y=2.0cm, scale=0.8, every node/.style={scale=0.8}]
\tikzstyle{var} = [draw, circle, minimum height=1cm,text centered, line width=0.5pt ]
\tikzstyle{obs} = [var]
\tikzstyle{train} = [var]
\node[var] at (0,0) (zt) {$z$};
\node[train, fill=blue!20] at (0.8, 0.8) (it) {$i$};
\node[obs, fill=blue!20] at (0, 0.8) (txt) {$x$};
\node[obs, fill=blue!20] at (0.8, 0) (rt) {$y$};
\draw[arrow] (zt) -- (it);
\draw[arrow] (zt) -- (txt);
\draw[arrow] (zt) -- (rt);
\end{tikzpicture}
\caption{\small Training}
\end{subfigure}
\quad\quad\quad
\begin{subfigure}{0.10\textwidth} %
	    \centering
	\begin{tikzpicture}[x = 2.0cm, y=2.0cm, scale=0.8, every node/.style={scale=0.8}]
\tikzstyle{var} = [draw, circle, minimum height=1cm,text centered, line width=0.5pt ]
\tikzstyle{obs} = [var]
\tikzstyle{train} = [var]
\node[var] at (0,0) (zt) {$z$};
\node[train] at (0.8, 0.8) (it) {$i$};
\node[obs, fill=blue!20] at (0, 0.8) (txt) {$x$};
\node[obs, fill=blue!20] at (0.8, 0) (rt) {$y$};
\draw[arrow] (zt) -- (it);
\draw[arrow] (zt) -- (txt);
\draw[arrow] (zt) -- (rt);
\end{tikzpicture}
\caption{\small Testing}
\end{subfigure}
\caption{Causal graphical model (CGM) for the training phase (a) and the testing phase (b).
The blue nodes $x$ and $y$ are observable (e.g., noisy speech and speech intelligibility scores). Node $z$, colored white, is not observable as a parameterized latent variable. Node $i$, colored blue in (a), is only observable during training, and to be inferred by proposed causal model in (b) the testing time.}
\label{fig:cgm}
\vspace{-0.4cm}
\end{figure}
Meanwhile, similar designs are sparse expert models~\cite{aswin2020sparse,kumatani2021building}. These approaches divide a large task into small sub-tasks by allocating data categorized in different attributes to several local expert models. Nevertheless, those models do not take into account the assumption of a causal graph; therefore, they can not be  evaluated under a formal causal learning settings with treatment effect analysis. Finally, causal effect measurements~\cite{pearl2019seven} could incorporate statistical refutation tests to design reliable prediction models.\par
This study focuses on develop a SE system using the potential outcome framework~\cite{rubin2005causal} shown in Fig.~\ref{fig:cgm}. Under this paradigm, we model the confounding variable $z$ that causally impacts SE performance $y$, instead of directly modeling the correlation between the noisy and clean speech.
To this end, a two-stage training procedure is adopted to attain satisfactory SE results by leveraging auxiliary intervention labels. To the best of our knowledge, the proposed training process and architecture are the first attempts to introduce CI into an end-to-end neural SE system. 
In contrast to previous studies that adopt prior causal features \cite{agrawal2015learning, vanwavenet} to improve system performance, we focus on the inference for vector-to-vector regression network of SE \cite{Qi2020slp, siniscalchi2021vector} by employing an auxiliary sub-task for state estimation, i.e., noise detection, which could leverage the advantages of causal inference. We design our observational inference enhanced network based on this causal neural architecture. Our contributions is fourfold: (i) a novel neural SE architecture based on causal inference aimed at handling complicated noisy condition is presented; (ii)  a novel quantitative measure for the causal effect of the selected intervention is devised; (iii) effectiveness of the proposed solution is demonstrated by showing that CISE outperforms both the non-causal counterpart model and other techniques leveraging complicated EMs in terms of both quality and intelligibility, and (iv) the proposed system merely uses 2.64\% of the computational time and 4.96\% of GPU memory as compared with the largest CISE variant.
\begin{figure*}[t!]
  \begin{subfigure}[b]{0.7\linewidth}
  \centering
  \includegraphics[width=\textwidth]{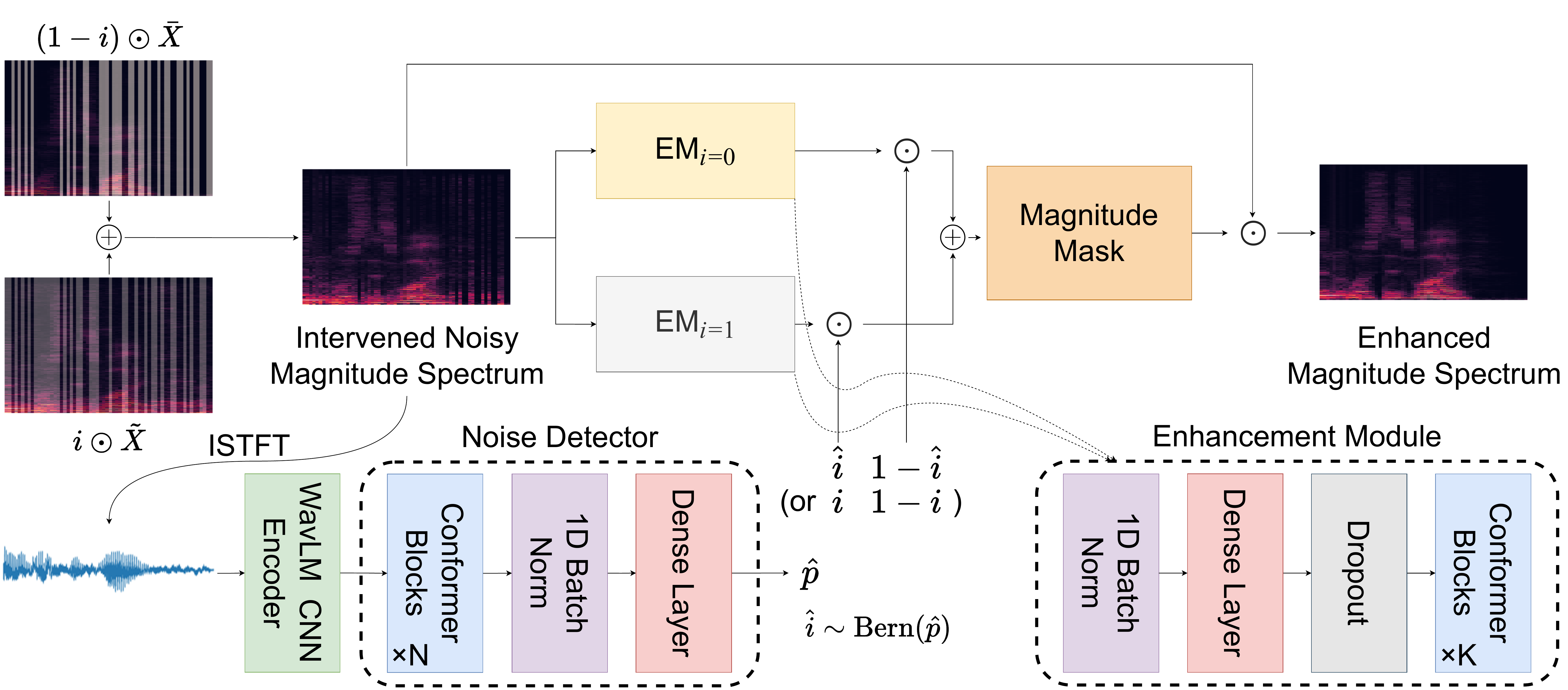}
  \caption{CISE training/testing flow.}
  \label{fig:system}
  \end{subfigure}
  \hfill
  \begin{subfigure}[b]{0.3\linewidth}
  \centering
  \includegraphics[height=5cm]{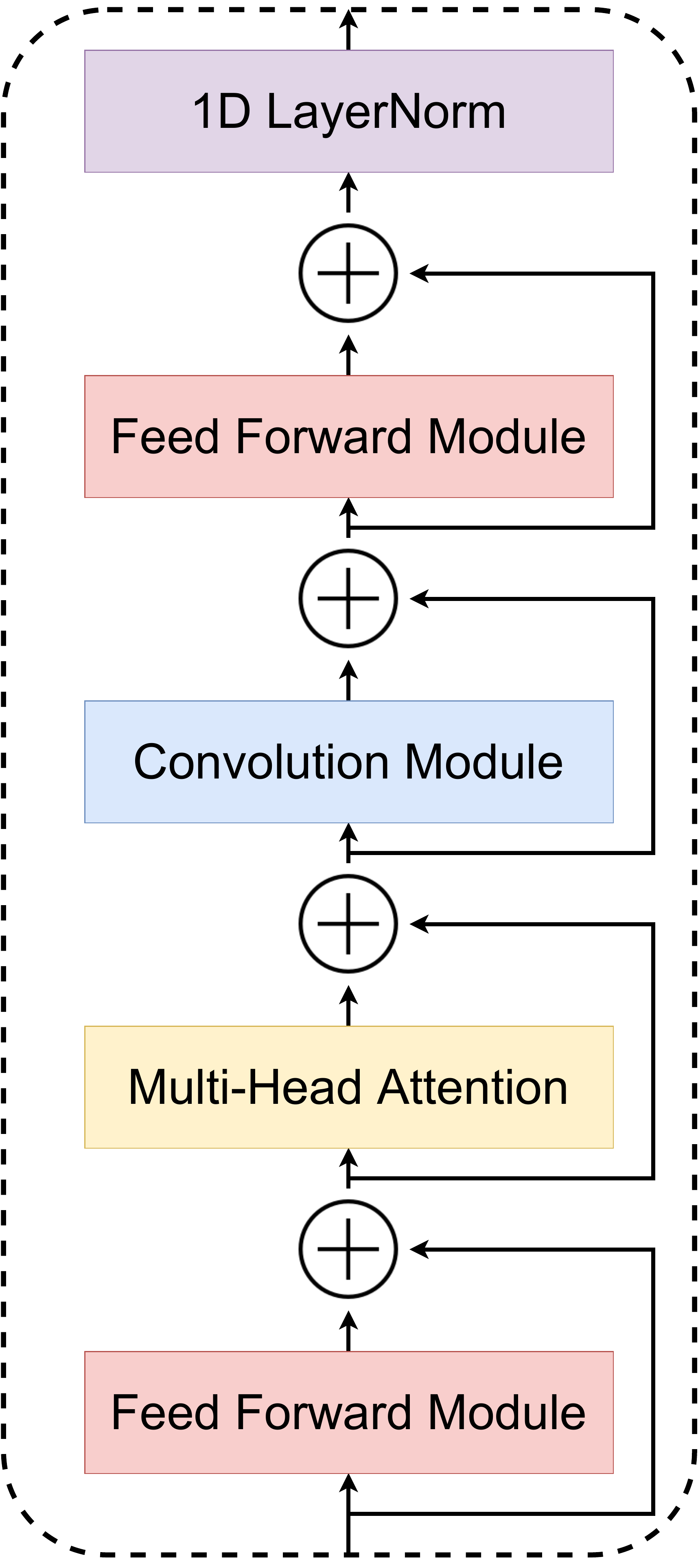}
  \caption{Conformer block.}
  \label{fig:conformer}
  \end{subfigure}
  \caption{Proposed causal inference-based SE architecture. Causal inference-based SE (CSIE) is based on two independent neural enhancement module (EM) without parameter sharing.  }
  \vspace{-0.4cm}
\end{figure*}
\section{Background}
\label{sec:bg}
\subsection{Causal Inference \& Representation Learning}
\label{sec:cidl}
Based on Pearl's causal hierarchy theorem~\cite{bareinboim2022pearl}, modeling machine learning problems at a higher causal level (e.g., interventional) could provide access to more useful information while extracting relevant features or in learning from proxy variables \cite{yang2022training}. %
Recently, causality learning~\cite{pearl2010causal, robins1995analysis} has been proven successful when combined with representation learning for feature extraction and probabilistic inference in sequence modeling \cite{melnychuk2022causal}. For instance, causal convolution~\cite{agrawal2015learning, vanwavenet} is a successful approach %
widely applied in speech synthesis. %
Moreover, observational inference empowered neural models attaining top performances in clinical learning~\cite{killian2022counterfactually}, sequence modelings \cite{wang2021inferbert}, and robust reinforcement learning~\cite{yang2022training}. %
Causal inference~\cite{pearl2010causal} is another mainstream approach to causal learning that focuses on learning robust proxy variables and inferencing under unseen dynamics, such as confounding variables.
\subsection{Treatment Effect Quantification}
\label{sec:ate}
To quantify the causal effects of the intervention on the outcomes of interest in a randomized controlled trial (RCT), the average treatment effect (ATE) \cite{holland1986statistics} is a metric often adopted. ATE measures the mean difference between the potential outcomes of the treatment and control groups and is formally defined as
\begin{align}
\label{eq:ate}
    \mathrm{ATE}=\mathbb{E}[y|i=1]-\mathbb{E}[y|i=0],
\end{align}
where $i$ is a binary label that indicates the occurrence of an intervention when its value is equal to one. The two terms on the right-hand side of Eq.~\eqref{eq:ate} denote the expected outcome over the population for the treatment and control groups, respectively. Briefly, a positive/negative ATE implies that the selected intervention has a positive or negative causal effect on the outcome of interest.

\section{Causal Inference-based SE system}

\subsection{Problem Definition}
\label{ssec:def}
In general, the causal graphical model of the potential outcome framework~\cite{rubin2005causal} can be represented as Fig.~\ref{fig:cgm}. In the CGM, $x$ is a noisy observation; $y$ is the outcome for that observation; $i$ denotes an intervention, and $z$ is a hidden confounding variable that affects the other three variables. Note that because $z$ is learnt implicitly, we cannot know the exact meaning of it. In training, $i$ is an observable variable used as a guidance to learn $z$ efficiently \cite{yang2022training}; however,  $i$ is unobservable and has to be predicted by the model at a testing time.

As we focus on forming the SE problem in the interventional causation, we integrate the SE problem with the potential outcome framework, as shown in Fig.~\ref{fig:cgm}. Here, we denote variables in the time–frequency domain in uppercase letters. For CISE, we intervene the clean speech by adding noise into it. Therefore, we define the intervened noisy speech $X=(1 - i) \odot \bar{X} + i \odot \tilde{X}$, where $\bar{X}$ and $\tilde{X}$ are the clean and noisy speech, and $i$ is a 0/1-mask that indicates in what frames the intervention occurs. In other words, the noise signal does not prevail over all time stamps. Fig.~\ref{fig:system} (a) illustrates how $i$ intervenes the clean speech. The outcome $y$ can be any measure of speech quality, intelligibility, or distance. As for CISE training, we set $y$ as the $l_1$ distortion distance. Therefore, CISE learns to model $z$ guided by a predefined intervention, which is further used to predict an intervention in the testing phase.

\subsection{Enhancement Modules and Noise Detector}
\label{ssec:cen}
The convolution-augmented Transformer (Conformer) \cite{gulati2020conformer} has been proven effective in various speech applications \cite{gulati2020conformer, miyazaki2020conformer}, including SE \cite{kataria2021perceptual, kim21h_interspeech}; therefore, we adopt it as a main component of our SE system. A Conformer consists of a series of half-step feed forward modules, a multi-head attention, and a convolution module as shown in  Figure \ref{fig:conformer}. We also employed  half-step feed forward layers \cite{lu2019understanding} and relative sinusoidal positional encoding for improved perfomance. %
In Fig.~\ref{fig:system}, two enhancement modules manage the processing of speech frames belonging to the treatment and control groups, respectively. Similar to \cite{miyazaki2020conformer}, each enhancement module consists of a batch normalization layer, linear transformation, followed by a few Conformer blocks. \par
In testing, we need to identify when an intervention occurs. Since  the presence of noise is regarded as an intervention in our study, the identification approach is implemented as a noise detector.
Although Mel-frequency cepstral coefficients (MFCCs) could  employed as input features for  noise detection, we observed that MFCCs led  to a severe overfitting of the training data, yielding a 36\% difference between the training and testing accuracy - domain mismatch may have caused that. To circumvent overfitting, we used WavLM~\cite{chen2021wavlm} CNN as an encoder to extract more generalizable embedding. Since noise presence prediction can be thought of as a sound event detection task, in which Conformers attained top accuracies~\cite{miyazaki2020conformer, wang2021four},  we used a Conformer-based noise detector. In the bottom left in Fig.~\ref{fig:system}, the noise detector is fed with embedding extracted by the WavLM module. The temporal dependencies among  embedding are then modeled by the Conformer blocks and mapped into a sequence of two-dimensional vectors, representing the probabilities of speech fragments being noisy or clean. Finally, the predicted intervention $\hat{i}$ is sampled from a Bernoulli distribution using the predicted probability $\hat{p}$.
\subsection{CISE Training and Testing}
\label{ssec:ma}
As shown in Fig. \ref{fig:system}, we take an intervened noisy speech magnitude spectrum, a staggered combination of the clean and the noisy speech frames in time, as the input. The intervention $i$, which guides the training procedure, is generated from a Bernoulli distribution (see Section \ref{sec:datacuration} for details); in testing, the noisy detector, shown in Fig. ~\ref{fig:system}(a), generates $\hat{i}$ in order to select different enhancement modules. 
Next, two Conformer-based enhancement modules, namely $\mathrm{EM}_{i=0}$ and $\mathrm{EM}_{i=1}$, which estimate magnitude masks belonging to the treatment ($i=1$) and control ($i=0$) groups, respectively, remix the predicted magnitude masks based on the intervention labels, and the remixed magnitude mask is multiplied
by the intervened noisy magnitude spectrum in an element-wise manner. The intervention labels (or predictions during testing) are then used for mixing the outputs of different enhancement modules. 
Finally, the SE process is accomplished by multiplying the intervened noisy magnitude spectrum by the magnitude mask. 

As CISE simultaneously learns to identify noise occurrences and to perform enhancement, we characterize CISE training as a multitask learning process, and thus we formulate the loss function as
\begin{align}
    \mathcal{L}_{total}(\bar{X}, \hat{X}, i, \hat{i}) \coloneqq \mathcal{L}_{l1}(\bar{X}, \hat{X}) + \mathcal{L}_{CE}(i, \hat{i}),
\end{align}
where $\hat{X}$ denotes enhanced speech. 
We select the $l_1$ distance and cross-entropy (CE) for the regression and classification tasks, respectively. For magnitude mask estimation, we simply minimize the $l_1$ distance between the enhanced and target spectra. Meanwhile, CISE training also minimizes the CE between the distribution of the predicted interventions $\hat{i}$ and that of the corresponding ground truth.

\begin{table}[t]
\centering
\caption{Speech qualities and intelligibility of CISE on the curated VoiceBank--DEMAND dataset. DA is the noise detection accuracy.}
\label{tab:cise_quality}
\begin{tabular*}{\linewidth}{l | c |P{7mm} P{7mm} P{7mm} P{7mm} P{7mm}}
\toprule
\toprule
    {\bf Model} &  {\bf DA} & {\bf PESQ} & {\bf CSIG} & {\bf CBAK} & {\bf COVL} & {\bf STOI} \\
    \midrule
    Oracle & 1.00 & 3.21 & 4.74 & 4.36 & 4.04 & 0.98 \\
    \midrule
    CISE & 0.92 & 3.15 & 4.70 & 4.27 & 3.98 & 0.97 \\
    CISE-C~\cite{cao22_interspeech}& 0.92 & 2.82 & 4.18 & 3.75 & 3.52 & 0.95 \\
    CISE-D~\cite{defossez2021hybrid} & 0.92 & 2.78 & 4.18 & 3.71 & 3.49 & 0.96 \\
    CISE-M~\cite{park2022manner} & 0.92 & 2.60 & 3.98 & 3.64 & 3.30 & 0.95 \\
    \midrule
    MFCC & 0.56 & 2.56 & 4.15 & 3.61 & 3.38 & 0.95 \\
    Random $\hat{i}$ & 0.50 & 2.50 & 4.02 & 3.50 & 3.28 & 0.94 \\
    Vanilla & -- & 2.44 & 3.68 & 3.20 & 3.05 & 0.93 \\
    $1-i$ & 0.00 & 2.23 & 3.57 & 3.05 & 2.89 & 0.92 \\
\bottomrule
\bottomrule
\end{tabular*}
\vspace{-0.4cm}
\end{table}
\subsection{ATE for Speech Enhancement}
For SE tasks, ATE can be defined for each chosen evaluation metric. Specifically, for a given metric $\mathcal{M}$ (e.g., PESQ or STOI), ATE is defined as
\begin{align}
\label{eq:atem}
    \mathrm{ATE}_{\mathcal{M}}=
    \mathbb{E}[\mathcal{M}(\bar{x}, \mathcal{E}(x))|i=1]-\mathbb{E}[\mathcal{M}(\bar{x}, \mathcal{E}(\bar{x}))|i=0].
\end{align}
In Equation~\eqref{eq:atem}, $\bar{x}$ denotes unobservable clean speech, $x$ is the observable noisy version of $\bar{x}$, and $\mathcal{E}(x)$ is the CISE-enhanced speech signal. \textcolor{black}{For example, a positive $\mathrm{ATE}_{\mathrm{PESQ}}$ implies that, on average, the addition of noise has a positive causal effect on the enhanced speech in terms of quality. The ATE is independent of optimization; therefore, CISE does not intentionally increase the ATE by distorting clean speech.}
\section{Experimental Setup \& Results}
\label{sec:exp}
\subsection{Data Curation}
\label{sec:datacuration}
To evaluate the proposed CISE approach, we use the Voice Bank--DEMAND dataset \cite{valentinibotinhao16_interspeech}, which we curate to make it suitable for causal inference. In the original Voice Bank--DEMAND, clean speech from 30 speakers are recorded in a studio room at sample rate of 48 KHz. Among those speakers, speech material from 28 speakers is used for training, and the rests are used for testing. The training set includes 10 types of noises added to the clean speech at 4 signal-to-noise-ratio (SNR) levels, ranging from 0 dB to 15 dB. For the test set, 5 unseen noises are added to the clean speech at SNR from 2.5 dB to 17.5 dB.\par
For CISE, we need to know where the intervention takes place. Therefore, an additional information is needed, namely a label indicating the presence of noise in a given speech frame. To this end, we remix noisy data through combining clean and noisy speech $X = (1 - i) \odot \bar{X} + i \odot \tilde{X}$
\noindent where $i\sim \mathrm{Bern}(0, 1),~ i\in \{0, 1\}$ indicates of the appearance of noise, and $i$ is drawn from a Bernoulli distribution with $p(i=1) = 0.5$. $X$ is the intervened noisy speech randomly mixed by the clean speech $\bar{X}$ and the original noisy speech $\tilde{X}$. With the intervened noisy speech $X$ and the corresponding $i$, a causal inference-based SE system can be implemented as shown in Fig.~\ref{fig:system}. Technically, we sample $i$ with the length of $X$, and then repeat each time stamp to the frequency dimension of $X$.
\begin{table}[tb]
\centering
\caption{Computational overheads. Each entry shows the time and memory usage of processing 1 second speech signal. The left and right of the slash shows the usage for batch size of 1 and 16. Both forward and backward pass are considered.}
\label{tab:ovh}
\begin{tabular*}{\linewidth}{l | P{9mm} P{11mm} P{11mm} P{11mm}}
\toprule
\toprule
     &   CISE & CISE-C & CISE-D & CISE-M \\
     \midrule
    CPU Time (s) & 2.20/2.39 & 2.50/49.79 & 1.79/53.19 & 2.56/50.46 \\
    GPU Time (s) & 0.03/0.12 & 0.142/4.54 & 0.08/3.90 & 0.10/4.02 \\
    GPU Mem. (GB)& 2.19/2.91 & 4.47/58.62 & 2.55/23.11 & 2.76/29.51 \\
\bottomrule
\bottomrule
\end{tabular*}
\vspace{-0.4 cm}
\end{table}
\begin{table*}[htb]
\centering
\caption{Causal inference explanation for speech intelligibility indexes. ATEs with controlled noise detection accuracy $p$. The left columns denote quality/intelligibility scores; on the right are the ATEs of the corresponding metrics.}
\label{tab:ctrl_p}
\adjustbox{max width=0.95\textwidth}{
\begin{tabular}{c | c c c c c c | c c c c c c}
\toprule
\toprule
    \multirow{2}{*}{\bf Accuracy $p$} & \multirow{2}{*}{\bf PESQ} & \multirow{2}{*}{\bf CSIG} & \multirow{2}{*}{\bf CBAK} & \multirow{2}{*}{\bf COVL} & \multirow{2}{*}{\bf STOI} & \multirow{2}{*}{\bf SSNR} & \multicolumn{6}{c}{\bf Average Treatment Effect} \\
    & & & & & & & {\bf PESQ} & {\bf CSIG} & {\bf CBAK} & {\bf COVL} & {\bf STOI} & {\bf SSNR} \\
    \midrule
    0.0 & 2.23 & 3.57 & 3.05 & 2.89 & 0.92 & 8.37 & -1.1872 & -0.8659 & -1.2546 & -1.1394 & -0.0477 & -9.819\\
    0.1 & 2.27 & 3.65 & 3.12 & 2.96 & 0.92 & 9.04 & -1.1275 & -0.7827 & -1.1659 & -1.0606 & -0.0423 & -9.0432\\
    0.3 & 2.37 & 3.83 & 3.29 & 3.11 & 0.93 & 10.58 & -0.9617 & -0.5979 & -0.9384 & -0.8662 & -0.0314 & -7.0541\\
    0.5 & 2.51 & 4.03 & 3.50 & 3.29 & 0.94 & 12.50 & -0.7247 & -0.3834 & -0.6063 & -0.6081 & -0.0193 & -3.9817\\
    0.7 & 2.69 & 4.26 & 3.77 & 3.51 & 0.95 & 14.95 & -0.3772 & -0.1548 & -0.1055 & -0.2598 & -0.0063 & 1.2719\\
    0.9 & 2.99 & 4.56 & 4.13 & 3.82 & 0.97 & 17.99 & 0.2283 & 0.0556 & 0.4185 & 0.2610 & 0.0099 & 10.0948\\
    1.0 & 3.21 & 4.74 & 4.36 & 4.04 & 0.98 & 19.83 & 0.9245 & 0.1013 & 0.4860 & 0.5576 & 0.0204 & 16.4458\\
\bottomrule
\bottomrule
\end{tabular}
}
\vspace{-0.2cm}
\end{table*}
\subsection{Experimental Setup}
To match the down sampling rate and the features size of the WavLM CNN encoder, we set the size of Fourier transform to 1023, the length of the analysis window to 1023 sample points (approximately 0.064 seconds), and the step of sliding window to 320. 
The dropout rate is set to 25\%, and we use two layers of Conformer blocks to encode temporal dependencies.
For intervention prediction, we use the WavLM CNN encoder to extract a 512-dimensional general purpose representation of the speech, which is then used for noise detection. The noise detector comprises four Conformer blocks stacked together, along with a batch normalization layer over the channels of features vectors and a fully-conntected layer to reduce the dimensionality of the hidden states from 512 to 2, representing $p(i=0)$ and $p(i=1)$, respectively. For convergence stability, in the training stage, we use the intervention labels for enhancement module switching; however, we only use the predicted interventions during testing. Finally, a standard Adam \cite{kingma2014adam} with learning rate $10^{-4}$, $\alpha=0.9$, and $\beta=0.999$ is adopted for the optimization.
\subsection{Speech Quality and Intelligibility Results}
\label{ssec:quali_intel}
In this section, we compare the proposed CISE system with its variants using the state-of-the-art SE models as EMs and analyze the importance of the noise detector. We report several metrics often used to assess the speech quality and intelligibility in Table~\ref{tab:cise_quality}. PESQ estimates the perceptual speech quality by assigning a score ranging from -0.5 to 4.5. The STOI is an intrusive measure of the intelligibility of degraded speech signals. CSIG, CBAK, and COVL are composite measures~\cite{loizou2008composite} of speech quality. CSIG focuses on the quality of foreground speech; conversely, CBAK estimates the extent of the intrusion of background noise, with a higher score indicating less intrusion; COVL evaluates overall quality combined with previous scores. SSNR represents for segmental signal-to-noise ratio. \par
In Table~\ref{tab:cise_quality}, from top to bottom, we report the speech quality and intelligibility scores of different SE systems. Oracle denotes the ideally achievable result with CISE when the intervention labels are known at the testing time. From the second to the fifth rows are CISE and its variants, where CISE, CISE-C, CISE-D, and CISE-M use Conformer-based (proposed), CMGAN~\cite{cao22_interspeech}, DEMUCS~\cite{defossez2021hybrid}, and MANNER~\cite{park2022manner} as EMs, respectively. Comparing these CISE variants, we observed that despite sharing the same detection accuracy, the results of using different EMs for CI training is uneven. CISE has overall highest scores; CISE-C and CISE-D have a similar behavior; CISE-M performs even worse than Random $\hat{i}$ (described in the following context) for CSIG and COVL. In addition, CISE with Conformer building blocks attains better results using less processing time, and with a lower memory consumption, as shown  in Table~\ref{tab:ovh}. Accordingly, we conclude that the proposed CISE with Conformer building blocks best matches the potential outcome framework among the studied settings. Starting from the sixth row, MFCC refers to a noise detector that uses MFCC features as inputs with identical classification structure as CISE presented in Section~\ref{ssec:cen}; Random $\hat{i}$ refers to a CISE architecture with a malfunctioning noise detector that samples $\hat{i}$ at random following a Bernoulli distribution as described in Section~\ref{sec:datacuration}. As expected, Random $\hat{i}$ achieves slightly worse results than MFCC since the detection accuracy of Random $\hat{i}$ is slightly higher than the detection accuracy of MFCC. In the eighth row, Vanilla leverages the same EM used by the CISE without the module switching mechanism; that is, only a single enhancement module is present. The last row, $1-i$, offers the worst case of the CISE system, where clean and noisy frames are swapped (i.e., the detection accuracy is 0), and thereby assigned to an incorrect enhancement module in Fig. \ref{fig:system}.
Through controlling the EM structure and comparing the CISE with the last four rows, we see that, CISE outperforms its non-causal version and other cases by taking advantage of being informed by the hidden factor behind observable variables, and thus CISE yields the best overall results. Notably, although the same Conformer-based SE architecture is used, Random $\hat{i}$ can still outperform Vanilla even with a randomized intervention.

\subsection{ATE with Speech Evaluation Metrics}
In Table~\ref{tab:ctrl_p}, we control the detection accuracy $p$ (shown in the left column) moving from 0.0 to 1.0. The speech quality and intelligibility scores discussed in Section~\ref{ssec:quali_intel} are reported in the second to the seventh columns; in the seventh column, the SSNR is presented. A CISE with higher detection accuracy leads to better results, demonstrating the importance of accurate intervention prediction.
The remaining columns in Table \ref{tab:ctrl_p} report the ATE values for each of the metrics given in the second to the sixth row. The ATE represents the causal effect of the selected intervention (i.e., the presence of noise) on an outcome (i.e., metric score). As each ATE monotonically increases as $p$ increases, the detection accuracy and ATE values are positively correlated. In addition, a $p$ greater than 0.9 is required to obtain positive ATEs in each metric, showing a high demand for an accurate noise detector for CISE inference.

\section{Conclusion}
\label{sec:con}
This study provided an effective and efficient solution to incorporate the neural SE training and inference with the potential outcome framework and consequently leveraged the power of causal inference. 
The experimental evidence showed that the proposed Conformer-based CISE system is capable of conducting outstanding performances regarding quality, intelligibility, and computational overheads with respect to computational time and memory usage.
Furthermore, CISE outperforms its non-causal counterpart and all the other variants using powerful state-of-the-art SE as EMs by a large margin, showing the imperative necessity in searching the suitable model for the EMs. Aside from the model performance, we also defined an ATE specialized for SE to quantify the causal effect on a metric given an intervention, and by investigating the change in ATE along with the manipulated detection accuracy, we argued that the accurate intervention prediction is crucial for inference since it causally impacted the performance of a SE system. Our implementation will be available at: \url{https://github.com/aleXiehta/Causal-SE}.

\clearpage
\bibliographystyle{IEEEbib}
\small
\bibliography{strings, refs, mybib}
\clearpage
\appendix
\section*{Appendix}
\subsection*{A. Causal Hierarchy and Speech Processing}
\begin{figure}[ht!]
  \centering
  \includegraphics[width=0.90\linewidth]{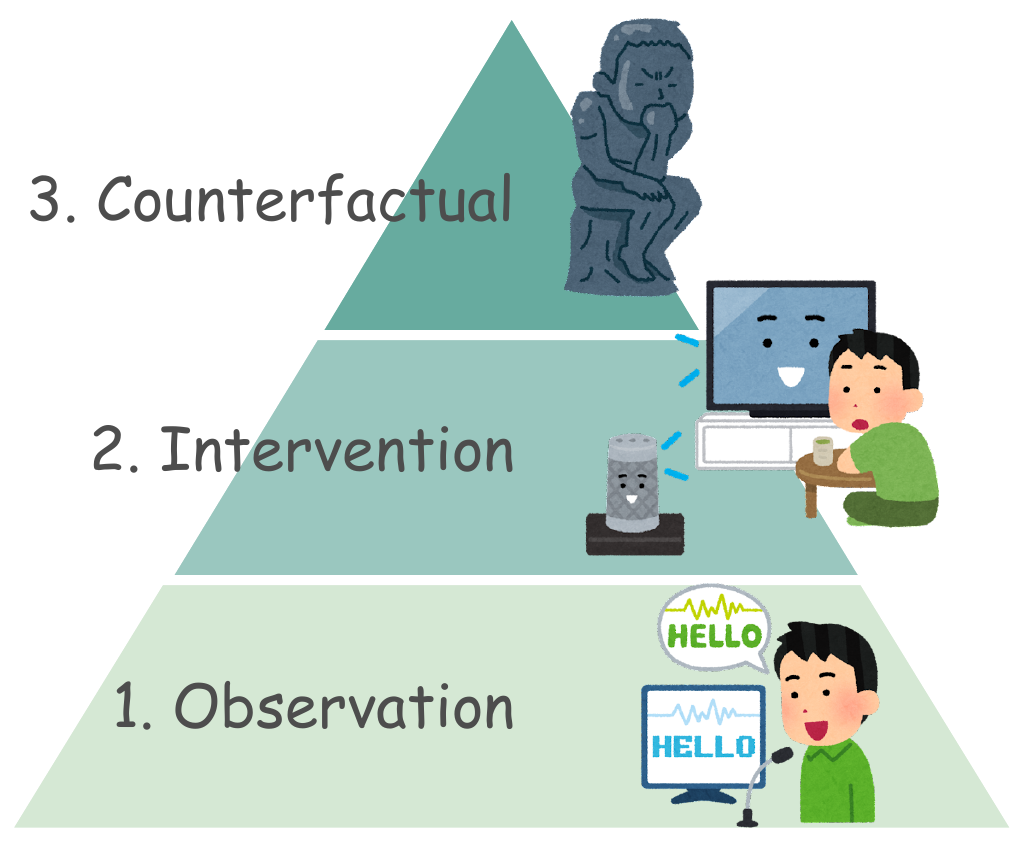}
  \caption{Automatic speech recognition (ASR) examples at each level of Pearl's causal hierarchy~\cite{bareinboim2022pearl}. \textcolor{black}{At the observation level, given an input “Hello,” an ASR algorithm recognizes speech by matching the input with the most likely word in its dictionary. Based on the first level, an ASR algorithm at the intervention level assumes that all observable variables are noisy and searches for a robust proxy variable representing words in the input speech. Given certain types of interventions, an ASR algorithm at the counterfactual level needs to imagine the following: ``\emph{what would the prediction be if different noise signals are involved?}'' Nevertheless, examples at this level require further investigation.}}
  \label{fig:PCH}
\end{figure}
Fig.~\ref{fig:PCH} depicts Pearl's causal hierarchy in the context of automatic speech recognition (ASR). At the \emph{observation level}, a non-causal ASR system is trained by maximizing the likelihood between the input speech and the corresponding labels (i.e., word index). At the \emph{intervention level}, the ASR system learns robust proxy variables from the intervened data, resulting in a better recognition performance when noise is involved. \textcolor{black}{At the \emph{counterfactual level}, one considers ``\emph{What would the outcome be if I took another action?}'' However, ASR applications at this level are yet difficult to deploy.}
Inferring this logic-aware information and adopting different effective strategies could be learned naturally through human perception. As a summary, how to design an end-to-end neural speech model equipped with the \textbf{power of ``inference''} is still an open topic and deserved more investigation.
For an SE system, it aims to improve speech quality and intelligibility of a speech signal. One major goal is to reduce noise contaminated speech. Given a clean speech signal $\hat{x}_t$ and a noise signal $n_t$ at time index $t$, a DL-based SE model tries to learn a mapping from noisy speech signal $x_t = \hat{x}_t + n_t$ to $\hat{x}_t$ so as to acquire clean speech. However, this formation could overly simplify the process of SE since, in many cases, noises are not prevailing in the whole utterance and likely unseen at test phase.

In the regime of causal learning, algorithms aim to model the hidden variable $z$, a confounder that obfuscates level 1 learning methods during testing, to achieve more accurate outcome prediction. In the scenario of SE, we define $x$ as noisy speech, $y$ as the likelihood of the enhanced and clean speech, and $z$ can be viewed as some concepts of clean speech that SE model wants to learn. By giving $i$, CISE can learn confounding variable $z$ more efficiently and robustly.
\subsection*{B. Algorithm of Causal Inference-based SE}
As presented in Algorithm~\ref{alg:lth}, the intervention label $i$ is used to train the enhancement modules and noise detector. During testing, the predicted intervention $\hat{i}$ is used to mix the estimated mask of each enhancement module. Therefore, the effectiveness significantly depends on the noise detector performance, and the modeling of the confounding variables.
\begin{algorithm}[tb!]
  \caption{Causal Inference-based SE system}
  \label{alg:lth}
\begin{algorithmic}
  \STATE {\bfseries 1.} \textbf{Inputs}: enhancement module, $\mathcal{G}$;  noise detector, $\mathcal{F}$; speech data, $\mathcal{D}$; \#Iteration, $K$; clean speech $\bar{X}$; original noisy speech $\tilde{X}$; intervened noisy speech $X$; enhanced speech $\hat{X}$; {\rm ISTFT} inverse short-time Fourier transform.\\
  \STATE {\bfseries 2.} \textbf{Randomly Initialize Weights}: $\theta_{i=0}^{(0)}$, $\theta_{i=1}^{(0)}$, and $\theta_{cls}^{(0)}$. \\
  \STATE {\bfseries 3.} \textbf{Training Iteration}: \\
  \FOR{$\bar{X}, \tilde{X} \sim \mathcal{D}$; $k^{th}$ iteration}
    \STATE $i \sim \mathrm{Bern}(p=0.5)$ %
    \STATE $X \gets (1 - i) \odot \bar{X} + i \odot \tilde{X}$ %
    \IF{training}
    \STATE $\hat{X} \gets (1 - i) \odot \mathcal{G}(X;\theta_{i=0}^{(n)}) + i \odot \mathcal{G}(X;\theta_{i=1}^{(n)})$
    \STATE $\hat{i} \gets \mathcal{F}({\rm ISTFT}(X); \theta_{cls}^{(n)})$
    \STATE $\theta_{i=0}^{(k+1)} \gets \theta_{i=0}^{(k)} - \gamma \nabla_{\theta_{i=0}^{(k)}} \mathcal{L}_{l1}(\bar{X}, \hat{X})$
    \STATE $\theta_{i=1}^{(k+1)} \gets \theta_{i=1}^{(k)} - \gamma \nabla_{\theta_{i=1}^{(k)}} \mathcal{L}_{l1}(\bar{X}, \hat{X})$
    \STATE $\theta_{cls}^{(k+1)} \gets \theta_{cls}^{(k)} - \gamma \nabla_{\theta_{cls}^{(k)}} \mathcal{L}_{CE}(i, \hat{i})$
    \ELSE
    \STATE $\hat{i} \gets \mathcal{F}(\mathrm{ISTFT}(x); \theta_{cls}^{(N)})$
    \STATE $\hat{X} \gets (1 - \hat{i}) \odot \mathcal{G}(X;\theta_{i=0}^{(N)})+ \hat{i} \odot \mathcal{G}(X;\theta_{i=1}^{(N)})$
    \ENDIF
    
  \ENDFOR

\end{algorithmic}
\end{algorithm}

\subsection*{Acknowledgement}
The authors want to thank insightful comments from Prof. Minje Kim, Indiana University Bloomington, on the preliminary draft. 
\end{document}